# Thermal excitation of plasmons for near-field thermophotovoltaics

Yu Guo, Sean Molesky, Huan Hu, Cristian L. Cortes and Zubin Jacob[*]

*Department of Electrical and Computer Engineering,*

*University of Alberta, Edmonton, Alberta T6G 2V4, Canada*

[*]*zjacob@ualberta.ca*

**Abstract:** The traditional approaches of exciting plasmons consist of using electrons (eg: electron energy loss spectroscopy) or light (Kretchman and Otto geometry) while more recently plasmons have been excited even by single photons. A different approach: thermal excitation of a plasmon resonance at high temperatures using alternate plasmonic media was proposed by S. Molesky et. al., *Opt. Exp.* 21.101, A96-A110, (2013). Here, we show how the long-standing search for a high temperature narrowband near-field emitter for thermophotovoltaics can be fulfilled by high temperature plasmonics. We also describe how to control Wein's displacement law in the near-field using high temperature epsilon-near-zero metamaterials. Finally, we show that our work opens up an interesting direction of research for the field of slow light: thermal emission control.

Nanoengineering the coherent state of light as in a SPASER[1] and the quantum state of light as in nanoscale single photon sources[2] have received significant attention over the last few years. However, controlling the thermal state of light i.e. nanoscale thermal sources with spectral tuning, narrow bandwidth and spatial coherence remains a challenge. In this paper, our aim is to show that thermally exciting plasmons[3] can lead to novel near-field thermal sources. This is a paradigm shift away from the conventional approaches of exciting bulk and surface plasmons through fast electrons[4], momentum matched light[5,6] or quantum emitters[7].

One major application of narrowband thermal sources with super-Planckian thermal emission can be in micron gap or near-field thermophotovoltaics[8–10]. The conventional Shockley-Quiesser limit of energy conversion only applies for a thermal source with the black body spectrum[11]. As opposed to this, if the thermal emission is spectrally matched to the bandgap of the cell, the efficiency of conversion can exceed the Shockley Queisser limit[12]. One important condition on the thermal source for thermophotovoltaics is the wavelength of emission which has to be between 1 μm < λ < 2 μm (T ≈ 1200 K) for compatibility with low bandgap thermophotovoltaic cells. We note that significant advances have been made in the field of tungsten photonic crystals[13–15] and metamaterials[16] for thermal engineering but not narrowband high temperature near-field emission. On the other hand, surface-phonon-polariton (SPhP) approaches for tailoring near-field thermal emission rely on natural material resonances (eg: SiC in the mid-IR) are not tunable to the spectral ranges relevant to low bandgap photovoltaics[17].

S. Molesky et. al. introduced the concept of high temperature plasmonics and metamaterials for far-field thermal emission control[18]. This required a switch away from conventional plasmonic building blocks like silver and gold to alternate plasmonic materials based on oxides and nitrides[19,20]. Note, conventional metals with plasma frequency at ultraviolet frequencies have shown thermal plasmonic effects[3] but would melt well before the black body temperature is able to provide efficient emission for thermophotovoltaic applications. We emphasize that alternate plasmonic media can have high melting points in the range of 3000 C allowing operation in the near-IR range crucial for practical applications. In this paper that we show that epsilon-near-zero[21] resonances and surface-plasmon-polaritons (SPP) using anisotropic multilayer plasmonic metamaterials can lead to narrowband super-Planckian thermal emission in the near-field. We also show that plasmonic slow light modes[22–24] behave like photonic van Hove singularities[25] that can lead to near-field thermal emission beyond the black body limit. We consider the potential application in near-field thermophotovoltaics where the narrowband thermal emission has to be between 1 µm < $\lambda$ < 2 µm for efficient energy conversion beyond the Shockley Queisser limit using low bandgap GaSb photovoltaic cells (E_gap = 0.7 ev). We emphasize that neither photonic crystals nor surface phonon polaritons can show narrowband super-Planckian near-field thermal emission in the near-infrared wavelength ranges.

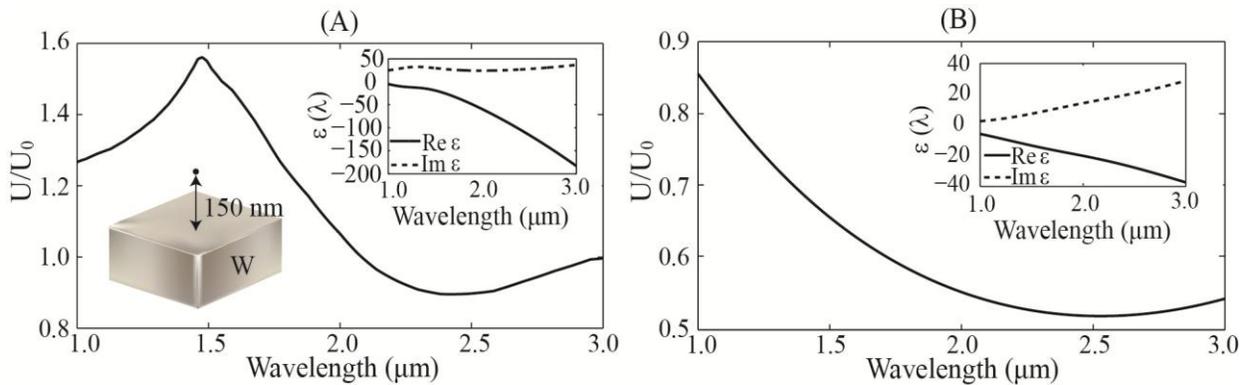

Fig 1(a) Normalized near-field thermal emission for a 500nm tungsten slab; the observation distance is 150nm. (b) Normalized near-field thermal emission for a 500nm TiN slab; the observation distance is 150nm. The dielectric constants are shown in the inset. Such broadband near-field sources are not suited for thermophotovoltaics.

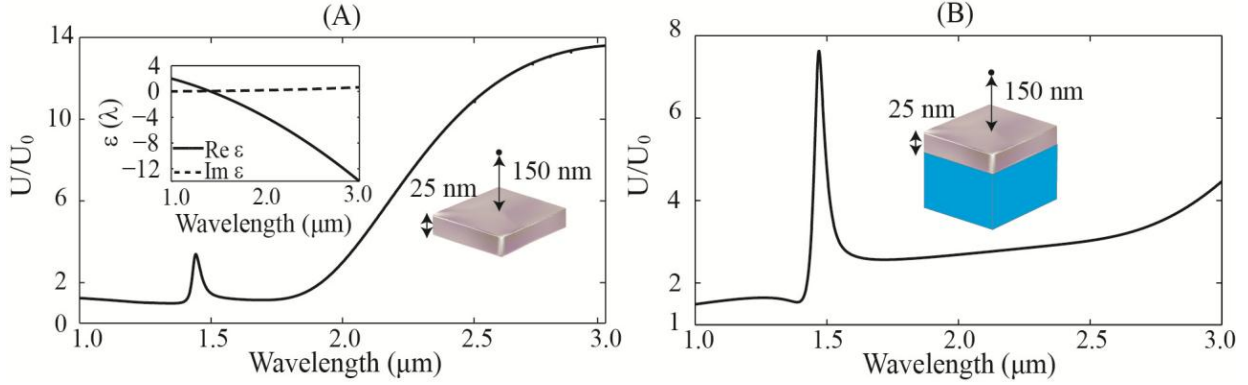

Fig. 2(a) Normalized thermal emission for a 25nm AZO slab; the observation distance is 150nm. Coupled surface plasmons of thin films play a key role in the thermal emission. (b) Normalized thermal emission for a 25nm AZO slab with a dielectric substrate ($\varepsilon=6$), the distance is 150nm. The substrate leads to tuning of the SPP resonance and a narrowband peak in the thermal emission.

Throughout the paper we use Rytov's fluctuational electrodynamics[26] to calculate the near-field thermal properties. The near-field energy density near a planar interface or slab is given by[27]

$$u(d,T) = \int_0^\infty d\lambda \frac{1}{2} u_0(\lambda,T) \sum_{j=s,p} \left\{ \int_0^1 \frac{k_\rho dk_\rho}{k_0 |k_z|} \frac{(1-|r_j|^2)}{2} + \int_1^\infty \frac{k_\rho^3 dk_\rho}{|k_z| k_0^3} e^{-2\operatorname{Im}(k_z)k_0 d} \operatorname{Im}(r_j) \right\}$$

where $u_0(\lambda,T) = 8\pi hc/(\lambda^5(e^{hc/\lambda k_B T}-1))$, T is the temperature of the body, d is the distance at which the energy density is measured, $r_s$ and $r_p$ are the reflection coefficients of (s) and (p) polarized waves incident on the interface, $k_\rho$ is the lateral wavevector and $k_z$ is the wavevector perpendicular to the interface (normalized to free space wavevector). We first consider the case of tungsten, a lossy metal with high temperature stability and analyze its' the near-field thermal emission (Fig. 1(a)). As is well known from Rytov's theory, any lossy medium possesses an enhanced non-radiative near-field local density of states $\rho^{n-rad}(d \ll \lambda) \approx \varepsilon''/(4|\varepsilon+1|^2 (k_0 d)^3)$ which is manifested in the near-field thermal energy density. These are also denoted by lossy

surface waves because such waves if incident on a surface are immediately absorbed. Note that high losses in the medium ($\varepsilon'' \gg 1$) can lead to a spectrally broad thermal emission beyond the black body limit however the effect is limited by the large absolute value of the dielectric constants $\rho^{n-rad}(d \ll \lambda) \propto 1/|\varepsilon+1|^2$. The goal therefore is to shift to metals with SPP resonance in the near-infrared range coupled with high temperature stability. In Fig. 1(b), we show the near-field thermal emission from Titanium Nitride, an alternate plasmonic metal[19] whose high temperature properties for thermophotovoltaics were pointed out by S. Molesky et. al[18]. The plasma frequency can be tuned by the deposition conditions however for recently reported values we see that the surface-plasmon-polariton peak cannot be efficiently excited at temperatures of 1000K. The corresponding figure insets show the stark contrast in the dielectric constant values of tungsten and Titanium nitride in the near-infrared range. The parameters are given in the supplementary information. Tuning the plasma frequency of Titanium Nitride to near-infrared ranges would make it an excellent candidate for thermophotovoltaic applications.

As opposed to this, aluminum doped zinc oxide (AZO) which has a plasma energy of 1.75 ev has a better enhancement in the near-field thermal emission (Fig. 2(a)). Beyond the epsilon-near-zero wavelength ($\lambda_{ENZ} \approx 1437$ nm), there is an enhancement in the near-field thermal emission from a thin film of AZO. The thin film supports two coupled surface plasmon polaritons (short range and long range) which gives rise to peaks in thermal emission. The broad peak is related to the short range surface-plasmon-polariton (SPP) and lossy surface wave contribution in the near-field of the thin film. This spectral behavior is not ideally suited for thermophotovoltaics. The goal of achieving a narrowband near-field thermal source can be achieved by using substrate effects to tune the SPP resonances. Fig. 2(b) shows the role of the substrate to tune the SPP resonance leading to a narrowband peak in thermal emission. Our analysis shows that the optimum solution for narrowband thermal sources consist of using simple thin films of high temperature plasmonic material with low damping and an appropriately chosen substrate for tuning.

We now discuss another route of achieving a narrowband super-Planckian thermal source with tunability. Multilayer metal-dielectric metamaterials also support epsilon-near-zero (ENZ) resonances and tunable anisotropic surface plasmon polariton resonances along with hyperbolic modes[28–30]. Previous work has shown that hyperbolic dispersion leads to broadband super-Planckian thermal emission[31,32]. We report here the narrow peak of thermal emission due to ENZ and anisotropic SPP modes which can be spectrally tuned by varying the fill fraction (Fig. 3). From Fig. 3(a), it is seen that as the fill fraction is changed from $\rho=1$ (thin metallic film) to $\rho=0$ (thin dielectric film), the physical origin of the thermal emission peak changes from isotropic surface waves to anisotropic surface waves and finally a bulk epsilon-near-zero mode at low fill fractions when $\rho<0.5$. We emphasize that thermally activated nonlinearities in the dielectric constituent of the metamaterial can also lead to similar tuning of the epsilon-near-zero wavelength but solely through temperature effects. This will correspond to a generalization of Wein's displacement law to the near-field.

Away from the epsilon-near-zero region multilayer thin film metamaterials show unique spectral signatures in the near-field thermal emission. In Fig. 4, we consider a multilayer metal-dielectric structure consisting of aluminum zinc oxide (AZO) and zinc oxide (ZnO). They achieve an effectively anisotropic response. The thermal excitation of modes of this structure leads to multiple narrowband peak in the near field (Fig. 4(a)). To understand the origin of the peaks, we plot the wavevector resolved local density of states[25] for the structure in Fig. 4 (b). The bright bands correspond to the modes of the structure which are thermally excited. It is seen that the near-field thermal energy density peak corresponds to the wavelength where the group velocity of the mode approaches zero.

This is a photonic van Hove singularity (VHS) that arises due to slow light modes in this coupled-plasmonic structure[25]. We argue that electronic equivalent of such thermal effects have been observed in lower dimensional systems like carbon nanotubes (CNTs). Quantum confinement is necessary to observe the enhancement in the electronic density of states due to the integrable nature of the singularity in 3D. Current driven thermal emission from CNTs has shown unique spectral features corresponding to electronic VHS[33]. Our results shown in Fig. 4 are the photonic equivalent of such phenomena. The tunneling of electrons measured using scanning tunneling microscopy (STM) exhibit an enhancement at energies corresponding to van Hove singularities in the carbon nanotube density of states[34]. The results of Fig. 4(a) can therefore be measured by the photonic equivalent of the STM experiment which is near-field thermal emission spectroscopy[27,35].

Another figure of merit of thermal sources along with narrowband operation is related to the spatial coherence[36]. Black body thermal emission is spatially incoherent as signified by the Lambertian radiation pattern in the far-field. The presence of gratings[37,38], photonic crystals[39] or epsilon-near-zero materials[18] can lead to thermal antenna effects i.e. spatial coherence in far-field thermal radiation. In the case of near-field thermal emission, the excitation of surface waves can lead to high degree of spatial coherence[40]. In Fig. 4(c), we analyze the near-field spatial coherence of our narrowband tunable thermal source arising from photonic van Hove singularities. We note that these are not surface states of the slab but are bulk waveguide modes. The spatial coherence is determined by the spectral density tensor $W_{ij}(r,r',\omega)$ which characterizes the correlation of the electromagnetic fields as a function of distance[40]. We see an excellent spatial coherence along the direction parallel to the planar interface due to the preferential thermal excitation of the slow light waveguide mode. This is discerned by the oscillatory nature of the field correlations as opposed to exponentially decaying nature expected from incoherent thermal emission from a black body. Similar effects are expected for the surface plasmon polariton waves as well.

A pertinent issue is the optimum design which achieves maximum heat transfer[41]. We emphasize that the present design from the previous section with a simple thin film supporting surface plasmon polaritons including substrate effects forms a promising solution. However, there can be exceptions depending on the temperature of operation and material properties. We calculate the near-field heat transfer[42] between thin film nanostructures at temperatures $T_1$ and $T_2$ separated by a gap d

$$H(d,T_1,T_2) = \int_0^\infty d\lambda (H_0(\lambda,T_1) - H_0(\lambda,T_2)) \sum_{j=s,p} \left\{ \int_0^1 k_\rho dk_\rho \frac{(1-|r_j^{01}|^2)(1-|r_j^{02}|^2)}{|1-r_j^{01} r_j^{02} e^{2ik_z k_0 d}|^2} + \int_1^\infty k_\rho dk_\rho e^{-2\mathrm{Im}(k_z)k_0 d} \frac{4\mathrm{Im}(r_j^{01})\mathrm{Im}(r_j^{02})}{|1-r_j^{01} r_j^{02} e^{2ik_z k_0 d}|^2} \right\}$$

where $H_0(\lambda,T) = 2\pi hc^2 / (\lambda^5 (e^{hc/\lambda k_B T} - 1))$ and other terms defined similar to the near-field energy density. We show that the near field heat transfer for our multilayer structures supporting the van Hove singularity can far exceed the value in tungsten. In Fig. 4(d), we plot the heat transfer between multilayer AZO/ZnO structures. The temperature of the two slabs are taken to be $T_1$=1500 K and $T_2$ = 300 K. We note the excellent agreement between effective medium predictions (EMT) of the location of the van Hove singularity and the practical multilayer structure (Transfer matrix method - TMM).

In conclusion, we have suggested a paradigm shift for plasmonics: thermal excitation. We have analyzed the potential of high temperature plasmonic metamaterials as narrowband tunable thermal sources for applications in thermophotovoltaics and near-field thermal stamping.

Currently material losses are a major impediment but future improvements in deposition techniques and optimization can lead to practical sources. Our work also paves the way for manipulating the near-field spatial coherence and bandwidth of thermal sources using metamaterials.

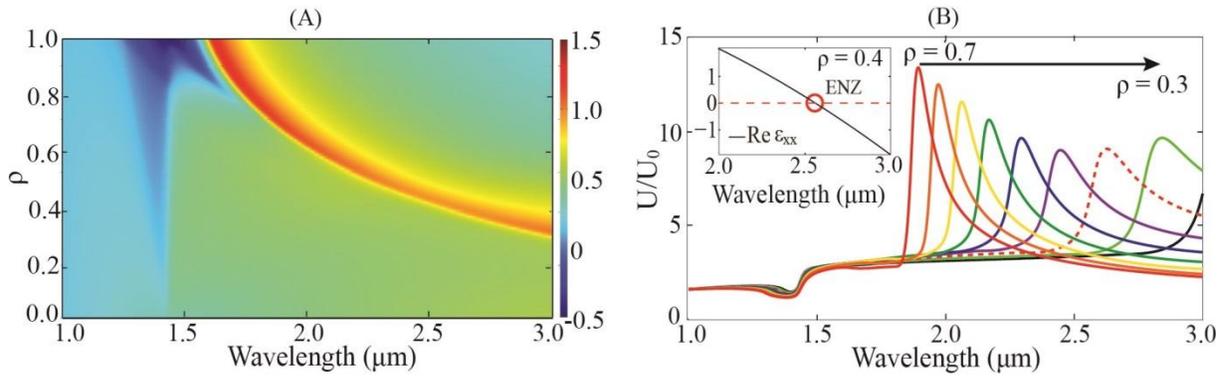

Fig.3 (a)Near field thermal emission (log scale plot) due to epsilon-near-zero resonances and anisotropic surface plasmon polaritons of multilayer AZO/ZnO; the observation distance is 150nm from the multilayer metamaterials. The bright red band is the super-planckian thermal emission which can be tuned by changing the fill fraction. (b) Thermal emission spectrum for fill fractions $\rho$ varying from 0.3 to 0.7. We can tune the thermal emission peaks within a wide spectrum. (right)

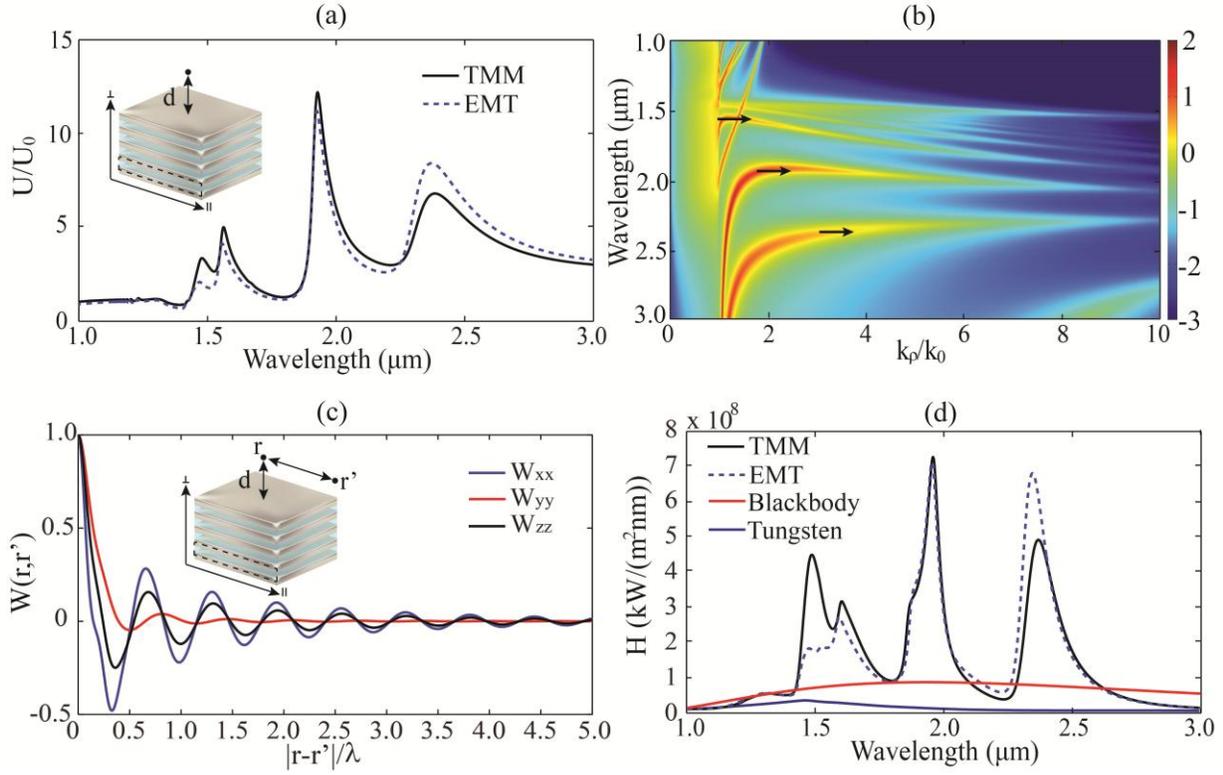

Fig.4 (a) Near-field thermal emission peaks due to slow light modes in multilayer structures consisting of Aluminum doped zinc oxide and Zinc oxide. These can be considered as photonic van Hove singularities where the density of states and thermal emission are enhanced. We have taken into account loss, dispersion, finite unit cell and finite sample size in our calculations. (Parameters: AZO and ZnO ($\varepsilon=6$) multilayers, each layer 25nm and 20 layers in total, the observation distance is 150nm.) (b) Wave-vector resolved thermal emission. The bright bands denote the modes of the structure which are thermally excited. We can clearly see slow light modes corresponding to the thermal emission peaks where the slope of the band goes to zero. (c) Spatial coherence at the wavelength (1927nm) which supports the slow light mode. Note that the thermal excitation of the slow light mode (coupled-plasmonic state) gives rise to spatial coherence in the near field. (d) Spectrally resolved near-field heat transfer between two identical multilayers with a gap of 150nm, $T_1$=1500K, $T_2$=300K. The black body heat transfer is shown for comparison. The slow light modes leads to super-Planckian heat transfer and there is an excellent correspondence between theory and the practical structures.

**Acknowledgements**

This work was partially supported by funding from Helmholtz Alberta Initiative, Alberta Innovates Technology Futures and National Science and Engineering Research Council of Canada.